\documentclass{blois}

\def\Journal#1#2#3#4{{#1} {\bf #2}, #3 (#4)}

\def\PLB{{\em Phys. Lett.}  B}
\def\PRL{\em Phys. Rev. Lett.}
\def\PRD{{\em Phys. Rev.} D}

\def\be{\begin{equation}}
\def\ee{\end{equation}}
\def\bea{\begin{eqnarray}}
\def\eea{\end{eqnarray}}

\newcommand{\Photo}{}
\usepackage{amsmath,amssymb}
\usepackage{bm}

\newcommand{\C}[2]{C_{\underset{#2}{#1}}}
\newcommand{\CL}[2]{\mathcal{C}_{\underset{#2}{#1}}}
\newcommand{\brackets}[1]{\left( #1 \right)}
\newcommand{\squarebrackets}[1]{\left[ #1 \right]} 

\begin{document}
\vspace*{4cm}
\title{FLAVOR ALIGNMENT OF NEW PHYSICS IN LIGHT OF \\
THE $\boldsymbol{(g-2)_\mu}$ ANOMALY}

\author{F. WILSCH }

\address{Physik-Institut, Universit\"at Z\"urich, CH-8057 Z\"urich, Switzerland}

\maketitle
\abstracts{
Addressing the ${(g-2)_\mu}$ anomaly and simultaneously being consistent with the tight bounds on ${\mu \to e \gamma}$ requires a precise alignment of the lepton dipole operator in flavor space. 
We investigate this alignment condition in the Standard Model Effective Field Theory~(SMEFT) by considering the Renormalization Group~(RG) mixing of the relevant operators. 
We find that semileptonic four-fermion operators, which are likely to yield a sizable contribution to the ${(g-2)_\mu}$ anomaly, need to be tightly aligned to the lepton dipole operator and Yukawa coupling. 
We discuss how New Physics~(NP) models can comply with these stringent conditions by employing particular dynamical assumptions or using flavor symmetries. 
}

\section{Introduction}
One of the longstanding puzzles in particle physics is the measurement of the muon magnetic moment~$a_\mu = (g-2)_\mu/2$, which was recently updated by the E989 experiment at FNAL.\cite{Muong-2:2021ojo} Combined with the previous BNL results\,\cite{Muong-2:2006rrc} the current discrepancy with the Standard Model~(SM) prediction\,\cite{Aoyama:2020ynm} is at $4.2\sigma$. This makes $a_\mu$ an excellent probe of physics beyond the SM~(BSM). 

Another indication for BSM physics is provided by the observation of lepton flavor non-universality in semileptonic $B$~decays.\cite{B-anomalies} At least in the case of the neutral-current $B$~anomalies these effects can also be explained by NP involving muons. It is therefore interesting to consider combined explanations of the $B$~anomalies and the $a_\mu$ anomaly as done for example in Refs.\cite{Greljo:2021xmg,Marzocca:2021azj,Arcadi:2021cwg} 

The purpose of this analysis\,\footnote{For further information on the analysis see Ref.\cite{Isidori:2021gqe} on which this article is based.} is to investigate the implications of the $a_\mu$ anomaly on the lepton flavor structure of the underlying NP models in general terms using the SMEFT. We furthermore discuss the compatibility of $a_\mu$ with the $B$~anomalies. Taken alone the $a_\mu$ anomaly does not imply any lepton flavor violating~(LFV) effects, however these can arise in models explaining the $B$~anomalies. We find that LFV must be strongly suppressed in combined explanations of these anomalies due to the tight constraints on~$\mu \to e \gamma$.
 
These proceedings are based on Ref.\cite{Isidori:2021gqe} and are organized as follows: in Sec.~\ref{sec:dipole} we discuss the lepton dipole operator in the SMEFT, its RG evolution, and the experimental constraints on its different components. We then find the flavor alignment conditions for all relevant operators required to satisfy these constraints. In Sec.~\ref{sec:alignment-mechanisms} we analyze how these alignment conditions can be achieved using either flavor symmetries or dynamical hypothesis.

\section{The leptonic dipole operator in the SMEFT}
\label{sec:dipole}
For the present analysis we assume that the NP responsible for the $a_\mu$ anomaly is heavy, i.e. far above the electroweak scale~$v\approx246\,\mathrm{GeV}$, such that its low-energy effects can be well described by the SMEFT considering effective operators up to dimension six. 
The Lagrangian containing both the lepton mass Yukawa coupling and dipole operator reads
\begin{align}
\mathcal{L}_\mathrm{broken} 
&\supset - [\mathcal{Y}_e]_{\alpha\beta} \frac{v}{\sqrt{2}} \left( \overline{e}_{L_\alpha} e_{R_\beta} \right)
+ \CL{e\gamma}{\alpha\beta} \frac{v}{\sqrt{2}} \left(\overline{e}_{L_\alpha} \sigma^{\mu\nu} e_{R_\beta}\right) F_{\mu\nu}
\end{align}
written here in the broken phase of the electroweak symmetry. Lepton flavor indices are indicated by $\alpha,\beta \in \{1,2\}$, where we only consider the two light generations for our analysis.

The dipole operator mixes with other SMEFT operators with the same lepton flavor structure under RG evolution. The relevant Lagrangian in the unbroken phase of the SMEFT is
\begin{align}
\begin{split}
\mathcal{L}_\mathrm{unbroken}  
\supset 
&- [Y_e]_{\alpha\beta} ( \overline{\ell}_{\alpha} e_{\beta} ) H
+ \C{eH}{\alpha\beta} ( \overline{\ell}_{\alpha} e_{\beta} ) H (H^\dagger H) 
\\
&+ \C{eB}{\alpha\beta} (\overline{\ell}_\alpha \sigma^{\mu\nu} e_\beta) H B_{\mu\nu} + \C{eW}{\alpha\beta} (\overline{\ell}_\alpha \sigma^{\mu\nu} e_\beta) \tau^I H W_{\mu\nu}^{I}
\\
&+ \C{lequ}{\alpha\beta ij}^{(3)} (\overline{\ell}_\alpha^n \sigma_{\mu\nu} e_\beta) \epsilon_{nm} (\overline{q}_i^m \sigma^{\mu\nu} u_j) 
+ \C{lequ}{\alpha\beta ij}^{(1)} (\overline{\ell}_\alpha^n e_\beta) \epsilon_{nm} (\overline{q}_i^m u_j)
+ \C{ledq}{\alpha\beta ij} (\overline{\ell}_\alpha^n e_\beta) (\overline{d}_i q_{jn}) \,
\end{split}
\end{align}
and the relations to the parameters of the broken phase are given by
\begin{align}
\label{eq:mass_Yukawa_broken_phase}
\begin{pmatrix}
	\CL{e\gamma}{\alpha\beta} \\ \CL{eZ}{\alpha\beta}
\end{pmatrix}
&= \begin{pmatrix}
	c_\theta & -s_\theta \\
	-s_\theta & -c_\theta
\end{pmatrix}
\begin{pmatrix}
	\C{eB}{\alpha\beta} \\ 
	\C{eW}{\alpha\beta}
\end{pmatrix} \,, 
&
\begin{pmatrix}
	[\mathcal{Y}_e]_{\alpha\beta} \\
	[\mathcal{Y}_{he}]_{\alpha\beta}
\end{pmatrix} 
&=
\begin{pmatrix}
	1 & -{1}/{2} \\
	1 & -{3}/{2}
\end{pmatrix}
\begin{pmatrix}
	[Y_e]_{\alpha\beta} \\
	v^2 \C{eH}{\alpha\beta}
\end{pmatrix} \, .
\end{align}

\subsection{Experimental constraints}
The muon anomalous magnetic moment~$\Delta a_\mu$ is determined by the combined results of the E989 experiment at FNAL\,\cite{Muong-2:2021ojo} and the E821 experiment at BNL\,\cite{Muong-2:2006rrc} using the SM prediction of Ref.\cite{Aoyama:2020ynm}
\begin{align}
\Delta a_\mu &= a_\mu^\mathrm{Exp} - a_\mu^\mathrm{SM} = (251 \pm 59) \times 10^{-11} \,,
& \text{with} &&
\Delta a_\mu &= \frac{4m_\mu}{e} \frac{v}{\sqrt{2}}\,\mathrm{Re}\, \CL{e\gamma}{22}^\prime
\label{eq:C_eg_22-bound}
\end{align}
relating (at tree-level) the anomalous magnetic moment to the lepton dipole Wilson coefficient with the prime indicating the mass eigenbasis of the charged leptons.

The branching ration for radiative LFV decays of leptons in the SMEFT (at tree-level) and the corresponding bound for the $\mu \to e \gamma$ decay as obtained by the MEG experiment\,\cite{MEG:2016leq} are
\begin{align}
\mathcal{B}(\ell_\alpha \to \ell_\beta \gamma) &= \frac{m_{\ell_\alpha}^3 v^2}{8 \pi \Gamma_{\ell_\alpha}} \left(|\CL{e\gamma}{\alpha\beta}'|^2 + |\CL{e\gamma}{\beta\alpha}'|^2\right) \, 
& \text{and} &&
\mathcal{B}(\mu^+ \to e^+ \gamma) &< 4.2 \times 10^{-13}~(90\%~\mathrm{C.L.}) \,.
\label{eq:C_eg_12-bound}
\end{align}

We can then use Eqs.~\eqref{eq:C_eg_22-bound} and~\eqref{eq:C_eg_12-bound} to find the constraints on the lepton dipole coefficients
\begin{align}
\mathrm{Re} \, \CL{e\gamma}{22}^\prime &\approx 1.0 \times 10^{-5} \, \mathrm{TeV}^{-2} \,,
&
|\CL{e\gamma}{12(21)}^\prime| <  2.1 \times 10^{-10} \, \mathrm{TeV}^{-2} \,.
\end{align}
These results can be combined in the flavor ratios
\begin{align}
|\epsilon^L_{12}|\,, ~ |\epsilon^R_{12}|  &<  2\times 10^{-5}\,,
&\text{where} &&
\epsilon^{L(R)}_{12} &\equiv { \CL{e\gamma}{12(21)}^\prime  }\,\Big/\,{    \CL{e\gamma}{22}^\prime    }\,.
\label{eq:eps-constraint}
\end{align}

\subsection{Renormalization group evolution}
The Yukawa coupling and the dipole operator mix with different Wilson coefficients as described by the RG equations $\mu \frac{\mathrm{d}}{\mathrm{d}\mu} C_k = \frac{1}{16\pi^2}\beta_k$, where the $\beta$-function of the SMEFT at $d=6$ have been derived in Refs.\cite{SMEFT:RGE}
Using these results we can express the dipole Wilson coefficient and the Yukawa matrix at the low scale~$\mu_L$ of the experiments in terms of the high NP scale~$\mu_H$ parameters
\begin{align}
\CL{e\gamma}{\alpha\beta} (\mu_L) 
&= \squarebrackets{1 - 3 \hat{L}  \brackets{y_t^2  + y_b^2}} \CL{e\gamma}{\alpha\beta} (\mu_H) - \squarebrackets{16\hat{L} y_t e} \C{lequ}{\alpha\beta 33}^{(3)} (\mu_H) \,,
\\
[\mathcal{Y}_e]_{\alpha\beta} (\mu_L) 
&= \left[ \squarebrackets{Y_{e}}_{\alpha\beta} - \frac{v^2}{2}\C{eH}{\alpha\beta} \right]_{\mu_H}
+ 6  v^2 \hat{L}
\left[ y_t^3 \C{lequ}{\alpha\beta 33}^{(1)}- y^3_b \C{ledq}{\alpha\beta 33} +  \frac34 (y_t^2  + y_b^2) \C{eH}{\alpha\beta} \right]_{\mu_H} \,,
\end{align}
where we neglect terms proportional to the quartic Higgs coupling or to at least two powers of gauge couplings and we define $\hat{L}=\frac{1}{16\pi^2}\log\frac{\mu_H}{\mu_L}$. Furthermore, we drop all terms proportional to Yukawa couplings other than top/bottom quark Yukawa~$y_{t/b}$.

\subsection{Flavor alignment conditions}
To use the experimental results discussed before, we must work in the mass eigenbasis of the charged leptons, i.e. we have to first diagonalize the lepton mass Yukawa coupling~$[\mathcal{Y}_e](\mu_L)$. Considering only the light ${2 \times 2}$~sector and small off-diagonal entries this is achieved by a rotation by $\Theta^{\mathcal{Y}}_{L(R)} = - \left. {[\mathcal{Y}_{e}]_{12(21)} }/{[\mathcal{Y}_{e}]_{22} }\right|_{\mu_L}$.
Applying the same rotation to the dipole operator we obtain $\CL{e\gamma}{12(21)}'(\mu_L) = \CL{e\gamma}{12(21)} (\mu_L) + \Theta_{L(R)}^{ \mathcal{Y}} ~\CL{e\gamma}{22} (\mu_L)$ and $\CL{e\gamma}{22}'(\mu_L) \approx \CL{e\gamma}{22} (\mu_L)$.

We can now define the left-handed \textit{flavor phases} for all relevant operator as the ratios of the 12 over 22 entries of the Wilson coefficients\,\footnote{The right-handed \textit{flavor phases} can be defined analogously as the ratio of 21 over 22 entries.} 
\begin{align}
\theta_{L}^Y  &= \left. \frac{ [Y_e]_{12}  }{   [Y_e]_{22} }  \right|_{\mu_H}   \,, &
\theta_{L}^{e\gamma}  &=  \left. \frac{ \CL{e\gamma}{12}   }{  \CL{e\gamma}{22}   } \right|_{\mu_H}    \,, &
\theta_{L}^{eH }  &=  \left. \frac{ \C{eH}{12}  }{ \C{eH}{22}  }  \right|_{\mu_H}  \,,  &
\theta_{L}^{u_i}  &= \left. \frac{ \C{lequ}{1233}^{(i)}   }{ \C{lequ}{2233}^{(i)}  }\right|_{\mu_H}   \,,  &
\theta_L^{d}   &=   \left.  \frac{ \C{ledq}{1233}  }{ \C{ledq}{2233}  } \right|_{\mu_H}  \,.
\label{eq:thetaX}
\end{align}
These phases are a measure of the flavor misalignment of the given operators at the high scale of~NP and do not depend on the absolute magnitude of the operators. We then find\,\footnote{Note that we express the dipole and Yukawa at the low scale whereas all other Wilson coefficients are at~$\mu_H$.}
\begin{align}
\begin{split}
\CL{e\gamma}{12}'(\mu_L) &=  (\theta_{L}^{e\gamma}-\theta_{L}^Y) \CL{e\gamma}{22}  (\mu_L) +    (\theta_{L}^{e\gamma}-\theta_{L}^{u_3}) (16 \hat{L} e y_t)  \C{lequ}{2233}^{(3)} (\mu_H)  \\
&\quad+  \left[   (\theta_{L}^Y-\theta_{L}^{u_1})  (6 y_t^3 ) \C{lequ}{2233}^{(1)}  (\mu_H)
+ (\theta_{L}^{d}-\theta_{L}^Y)  (6 y_b^3 ) \C{ledq}{2233} (\mu_H) \right]   \frac{ 1 }{  y_\mu }  \hat{L}  v^2  \CL{e\gamma}{22} (\mu_L)
    \\ 
&\quad+   (\theta_{L}^{eH } -\theta_{L}^Y)    \frac{1 -9( y_t^2 +y_b^2) \hat{L}  }{2} \C{eH}{22}  (\mu_H)   \frac{ 1 }{  y_\mu }   v^2  \CL{e\gamma}{22} (\mu_L)~.
\label{eq:Ceg-master}
\end{split}
\end{align}
The above equation allows us to derive a few important considerations:
\textit{i}) Even if~$\theta_L^{e\gamma}=0$, a non-vanishing $\CL{e\gamma}{12}'(\mu_L)$ is naturally generated at the low scale if any of the other phases $\theta_L^{X}$ in Eq.~(\ref{eq:thetaX}) is non-zero.
\textit{ii}) $\CL{e\gamma}{12}'(\mu_L)=0$ is obtained only aligning \textit{all} the flavour phases in Eq.~(\ref{eq:thetaX}).
\textit{iii}) All terms but one in Eq.~(\ref{eq:Ceg-master}) are proportional to the 22 entry of the dipole operator, which is required to be sizeable by the $a_\mu$ anomaly. 
\textit{iv}) The term proportional to $C_{ledq}$ has a numerically negligible coefficient and thus no significant tuning on the phase difference ${(\theta_L^d - \theta_L^Y)}$ is implied.
\textit{v}) The phase difference ${(\theta_L^{eH} - \theta_L^Y)}$ controls the flavor violating couplings of the Higgs boson, which are tightly constrained\,\cite{CeH} and can thus be neglected.
We can then conclude that a tuning of three phases differences is required to satisfy the constraints in Eq.~\eqref{eq:eps-constraint}:
\begin{align}
\epsilon^L_{12} &= 
(\theta_{L}^{e\gamma}-\theta_{L}^Y) +
(\theta_{L}^{u_3}- \theta_{L}^{e\gamma})  \Delta_{3} + 
(\theta_{L}^{u_1} - \theta_{L}^Y)   \Delta_{1} \,,
\label{eq:main12}
\end{align}
where we expect\,\footnote{For the definitions of $\Delta_{1/3}$ see Ref.\cite{Isidori:2021gqe} The estimate $\Delta_3 = \mathcal{O}(1)$ holds if $\Delta a_\mu$ is dominantly induced by $C_{lequ}^{(3)}$.} $\Delta_3 = \mathcal{O}(1)$ and $\Delta_1 = \mathcal{O}(10^{-3})$ compared to $\epsilon_{12}^L \lesssim 10^{-5}$.

\section{Flavor alignment mechanisms}
\label{sec:alignment-mechanisms}
The alignment conditions presented in Eq.~\eqref{eq:main12} could be satisfied accidentally in a specific NP model. Contrary to such tuned solutions, we can satisfy the given constraints on general grounds either employing some dynamical assumptions or using flavor symmetries.

\subsection{Dynamical alignment}
A natural choice to fulfill the condition in Eq.~\eqref{eq:main12} is to employ three dynamical assumptions:
\textit{I}) The dipole operator is not directly generated by the NP, but only induced by $C_{lequ}^{(3)}$ via RG mixing, leading to $\theta_L^{e\gamma}=\theta_L^{u_3}$. 
\textit{II}) The Wilson coefficients $C_{lequ}^{(1)}$ and $C_{lequ}^{(3)}$ are induced by the same NP dynamics and therefore have the same flavor structure, i.e. we have $\theta_L^{u_1}=\theta_L^{u_3}$. This is for example the case for the exchange of a single scalar leptoquark field.
\textit{III}) The muon Yukawa coupling is radiatively generated by the same NP dynamics as $C_{lequ}^{(1)}$ leading to $\theta_L^{Y}=\theta_L^{u_1}$. This condition is more challenging as the Yukawa is a marginal operator. Thus additional symmetries forbidding the Yukawa at tree-level are required to achieve this alignment. A model satisfying conditions \textit{I}-\textit{III} then automatically agrees with the tight constraints imposed by Eq.~\eqref{eq:main12}.

\subsection{Flavor symmetries}
The $U(2)^2=U(2)_L \otimes U(2)_E$ flavor symmetry acting on the two light generations is frequently used in the context of the $B$~anomalies.\cite{U2} It incorporates the hierarchical structure of the Yukawa matrices and allows for TeV-scale NP coupled dominantly to the third generation, while being consistent with the tight bounds from flavor violating processes. In the minimal breaking scenario the Yukawa can be parametrized in terms of two spurions~$V_\ell=(2,1)$ and~$\Delta_e=(2,\bar{2})$~by
\begin{align}
Y_e &= y_\tau
\begin{pmatrix}
	\Delta_e &   V_\ell \\
	0 & 1
\end{pmatrix} \,,
&
V_{\ell} &= 
\begin{pmatrix}
	0 \\ \epsilon_{\ell}
\end{pmatrix} \,,
& 
\Delta_e &= O_e^\intercal
\begin{pmatrix}
	\delta^\prime_e & 0 \\
	0 & \delta_e
\end{pmatrix}
\end{align}
with  $|\delta_e^\prime|  \ll  |\delta_e|  \ll  |\epsilon_\ell| \ll 1$ and 
where $O_e$ is a real orthogonal matrix ($[O_e]_{12} = \sin \theta_e \equiv s_e$). 

Performing a spurion expansion\,\cite{Faroughy:2020ina} $X_{\alpha\beta}^n = a_n (\Delta_e)_{\alpha\beta} + b_n (V_\ell)_\alpha (V_\ell^\dagger)_\gamma (\Delta_e)_{\gamma\beta}$ for all EFT operators we can express the \textit{flavor phases} as $	\theta_L^k \approx \frac{s_e}{c_e} (1 - \frac{b_k}{a_k}\epsilon_\ell^2)$, where $a_k,b_k$ are $\mathcal{O}(1)$ numbers.\footnote{At leading order in the spurion expansion no misalignement is generated, thus we consider subleading terms.} We can now obtain the constraint on the  phase difference of dipole and Yukawa operator
\begin{align}
	\left|  \theta_{L}^{e\gamma}-\theta_{L}^Y  \right|   =\left| \frac{s_e}{c_e}  \right| \epsilon^2_\ell \left|\frac{b_Y}{a_Y} - \frac{b_{e\gamma}}{a_{e\gamma}}\right| \leq 2 \times 10^{-5}.
\end{align}
With the assumptions\,\cite{U2-motivation} $\epsilon_\ell = \mathcal{O}(10^{-1})$, $c_e=\mathcal{O}(1)$, and $s_e=\mathcal{O}(\sqrt{m_e/m_\mu})$ this leads to the condition ${|s_e(\frac{b_Y}{a_Y} - \frac{b_{e\gamma}}{a_{e\gamma}})|} \lesssim 10^{-3}$ implying a tuning on the coefficients $a_k$, $b_k$ assumed to be of order~$\mathcal{O}(1)$. Thus the $U(2)$ symmetry is not enough to force the alignment required by Eq.~\eqref{eq:main12} on its own and needs further ingredients or tuning. 

A more efficient alternative is individual lepton flavor number conservation, i.e. employing $U(1)$ symmetries for the different lepton flavors. Assuming such symmetries to be good symmetries of the NP model all the flavor phases in Eq.~\eqref{eq:thetaX} vanish (see e.g. Ref.\cite{Greljo:2021xmg} for an example).

\section{Conclusion}
In the SM the flavor of charged leptons is conserved, but this symmetry can in principle be violated at higher energies by NP. In this work we have considered the lepton dipole operator which mediates both the $a_\mu$ anomaly as well as $\mu \to e \gamma$. Considering $\Delta a_\mu$ to be caused by NP sets a reference scale for the dipole operator which then has a precise orientation in flavor space due to the non-observation of $\mu \to e \gamma$. A set of other effective operators then needs to be tightly aligned to the dipole and the Yukawa as these operators mix under RG evolution and would, if not aligned, induce a too large LFV contribution in the dipole. These constraints on the flavor phases of the different operators are shown in our main result Eq.~\eqref{eq:main12}.
We can conclude that if $\Delta a_\mu$ is confirmed to be NP, the lepton sector must feature enhanced symmetries, in particular the conservation of the electron flavor, even beyond the SM. These properties are however not shared by the quark sector, making them behave quite differently at higher energies.


\section*{Acknowledgments}
I am grateful to my collaborators Gino Isidori and Julie Pag\`es. 
This project has received funding from the European Research Council (ERC) under the European Union's Horizon 2020 research and innovation programme under grant agreement 833280 (FLAY), and by the Swiss National Science Foundation (SNF) under contract 200020-204428.

\end{document}